\documentclass[12pt]{article}
\usepackage[utf8]{inputenc}
\usepackage[english]{babel}
\usepackage{amsmath}
\usepackage{amsfonts}
\usepackage{amssymb}
\usepackage{graphicx}
\usepackage{pgf,tikz}
\usepackage{mathrsfs}
\usetikzlibrary{arrows}

\author{G.~Chadzitaskos}
\title{On the influence of rings on orbital velocities}
\begin{document}

\maketitle

\abstract{ 
We analyze the possible effect of rings on orbital velocities in galaxies. The superposition of the central force with the gravitational forces induced by the rings opens up various possibilities of the course of orbital velocities. The orbital velocity depends on the position of the star in the ring. We illustrate this dependence on several models, where we show the course of potential curves and the curves of field strength.}

 \section{Introduction}
 
The aim of this work is to show that the problem of orbital velocities that do not decrease with distance from the center of galaxies can be partially explained by a ring effect. We demonstrate the course of the gravity potential and the course of the central forces in the case of a symmetric five rings galaxy (cylindrical symmetry). It is shown that there are places where the contributions of forces from the rings are positive, there are places where they are negative, and there are places where they are zero. 

For a distribution of orbital velocities in galaxies, see \cite{Dark}. The list of ring galaxies is in \cite{B3}

Similar potential and field curves can be expected in spiral galaxies, where a similar effect works, espetially  when the difference against ring galaxies is not too large, see \cite{B4}

The question arises as to whether it is possible to retrospectively study star formation and the subsequent composition of matter in individual parts of galaxies using the ring effect.

The idea is based on the different course of the gravitational potential for the sphere and the ring. The forces of the gravitational fields are substantially different in both cases.

  While the potential inside the homogeneous sphere is constant (and the field is zero), the negative potential of the ring decreases from the center of the ring to the inner circumference of the ring and increases with increasing distance from the outer circumference. Inside the ring there is a circle where the force is zero.

In the case of a homogeneous ball, the potential from the center increases with the square of the distance to the surface of the ball and then increases inversely proportional to the distance from the center.

\section{Gravitational potential and  field strength}

The gravitational potential $\phi (\vec{r})$ of mass in volume $ V $  at  point $\vec r$  is  
$$
\phi (\vec r) = - \int  \int \int_{V} \frac{G\,\, \rho (\vec R) }{|\vec r - \vec R|} dV,  $$ 
where $\rho (\vec R)$ is mass density.

The field strength is a negative potential gradient
$$\vec E(\vec r) = - \triangledown \phi (\vec r),$$
and the orbital speed for circular motion in symmetric field is
$$\frac{v^2}{r} = \vec E(\vec r) = - \triangledown \phi (\vec r). $$

\section{Potential of ring of finite height}

We express the course of the potential of a homogeneous ring with an inner diameter of $R_-= 800$, an outer diameter of $R_+=900$ and a height of $h=10,$ in cylindrical coordinates $R\in (800,900), \, z\in (-5,5),$ see Fig.1.
\begin{figure}[h]\label{Ring}
\includegraphics[scale=.8]{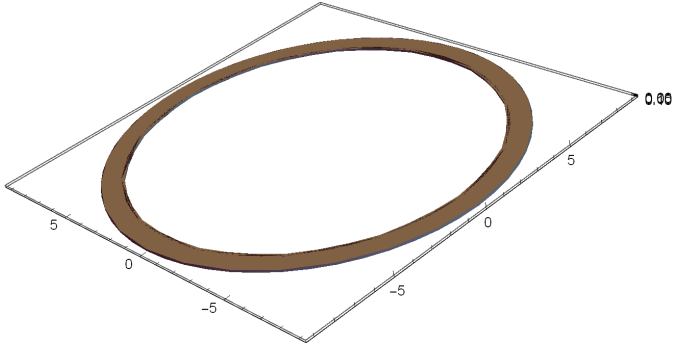}
\caption{Ring with inner diameter of $R_-= 800$, an outer diameter of $R_+=900$ and a height of $h=10,$}
\end{figure}

The gravitational potential of such a ring at a distance $r$ from the center in the plane $z = 0$ is given by the integral

$$
\phi (r) = - \int_{R_-} ^{R_+} G\,\, \rho  \int_0^{2 \pi} \int_{-h/2} ^{h/2} \frac{R \, \,  d \theta \,\,d z \,\, dR }{\sqrt{z^2 + R^2 + r^2 - 2 r R \cos \theta}}.  $$ 

In order to show the courses we can take values $G\, \rho= 1$ for a homogeneous ring 

With the help of numerical integration of the Mathematica  program we get (up to a constant multiple).
$$
\text{NIntegrate}
\frac{1}{1000}\left[\frac{R}{\sqrt{r^2-2 r R \cos (y)+R^2+z^2}},\{y,0,2 \pi \}, \{z,0.05,5\},\{R,800,900\}\right]$$

The course of potential and field strength is shown in Fig.2 and Fig.3.

  \begin{figure}[hbt!]\label{F0}
  \includegraphics[scale=.8]{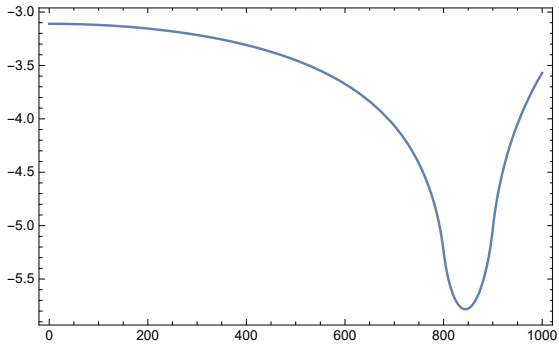}
  \caption{Gravitational potential $\phi (r)$ of the homogeneous ring}
 \end{figure}
 
\begin{figure}[hbt!]\label{F0F}
  \includegraphics[scale=.8]{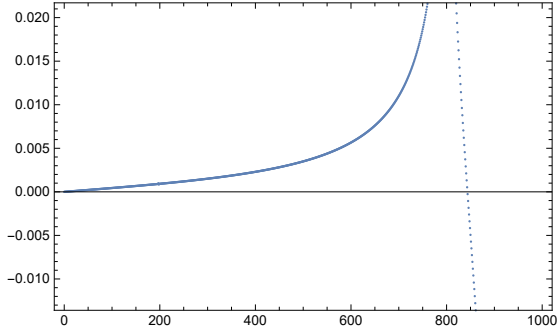}
  \caption{Gravitational field of the homogeneous ring}
 \end{figure}

 \section{Five rings model}

Similarly, we express the course of potential and field for a model of homogeneous rings, all of them of the same thickness and the same width, Fig. 4.
\begin{figure}[hbt!]\label{5R}
\includegraphics[scale=.8]{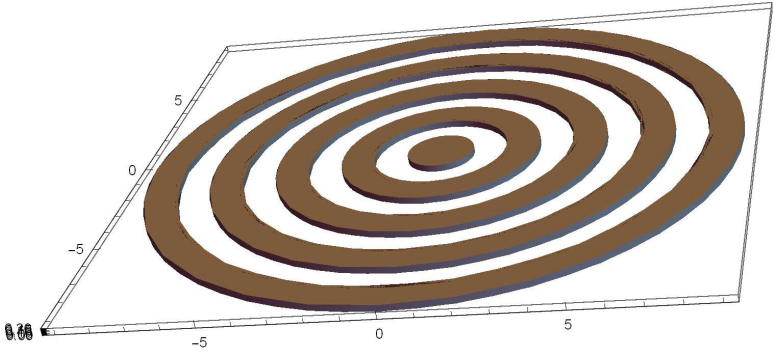}
\caption{Five rings model}
\end{figure}

The numerical calculation

$
\phi(r)=-\frac{1}{1000}  (\text{NIntegrate} [\frac{R}{\sqrt{r^2-2 r R \cos (y)+R^2+z^2}},\{y,0,2 \pi \},\{z,0.05,5\},\{R,0,100\}] \\+\text{NIntegrate}\left[\frac{R}{\sqrt{r^2-2 r R \cos (y)+R^2+z^2}},\{y,0,2 \pi \},\{z,0.05,5\},\{R,200,300\}\right] \\+\text{NIntegrate}\left[\frac{R}{\sqrt{r^2-2 r R \cos (y)+R^2+z^2}},\{y,0,2 \pi \},\{z,0.05,5\},\{R,400,500\}\right]\\+\text{NIntegrate}\left[\frac{R}{\sqrt{r^2-2 r R \cos (y)+R^2+z^2}},\{y,0,2 \pi \},\{z,0.05,5\},\{R,600,700\}\right]\\+\text{NIntegrate}\left[\frac{R}{\sqrt{r^2-2 r R \cos (y)+R^2+z^2}},\{y,0,2 \pi \},\{z,0.05,5\},\{R,800,900\}\right])$
gives the course of the potential in Fig. 5 and the course of the field strength in Fig. 6.

 \begin{figure}[hbt!]\label{F1}
  \includegraphics[scale=.8]{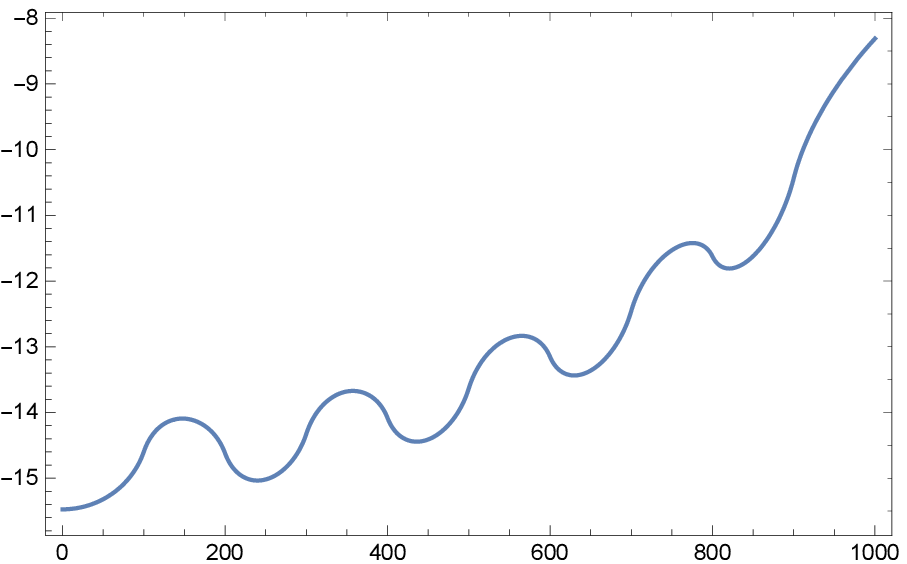}
  \caption{Gravitational potential $\phi (r)$ of the homogeneous five ring model}
 \end{figure}

\begin{figure}[hbt!] \label{F1F}
  \includegraphics[scale=.8]{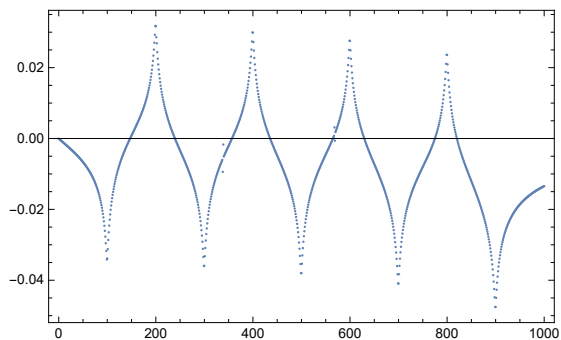}
  \caption{Gravitational field of the homogeneous five ring model}
 \end{figure}
   
The rings themselves act as attractors and the resulting force is a superposition of force from the central mass and the force caused by the rings in a galaxy. 
 \subsection{Decreasing mass density of rings}
The case where the density decreases as $1/R$ is given below

$
\phi(r)=-\frac{1}{1000}  (\text{NIntegrate} [\frac{1}{\sqrt{r^2-2 r R \cos (y)+R^2+z^2}},\{y,0,2 \pi \},\{z,0.05,5\},\{R,0,100\}] \\+\text{NIntegrate}\left[\frac{1}{\sqrt{r^2-2 r R \cos (y)+R^2+z^2}},\{y,0,2 \pi \},\{z,0.05,5\},\{R,200,300\}\right] \\+\text{NIntegrate}\left[\frac{1}{\sqrt{r^2-2 r R \cos (y)+R^2+z^2}},\{y,0,2 \pi \},\{z,0.05,5\},\{R,400,500\}\right]\\+\text{NIntegrate}\left[\frac{1}{\sqrt{r^2-2 r R \cos (y)+R^2+z^2}},\{y,0,2 \pi \},\{z,0.05,5\},\{R,600,700\}\right]\\+\text{NIntegrate}\left[\frac{1}{\sqrt{r^2-2 r R \cos (y)+R^2+z^2}},\{y,0,2 \pi \},\{z,0.05,5\},\{R,800,900\}\right])$

The resulting courses are shown in Fig. 7 and Fig. 8.

 \begin{figure}[hbt!]
  \includegraphics[scale=.8]{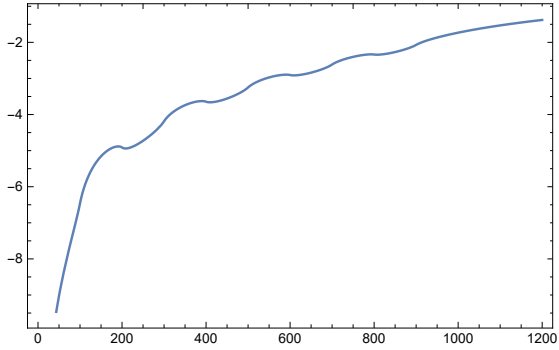}
  \caption{Gravitational potential $\phi (r)$ of the five ring model with $1/R$ density}
 \end{figure}

\begin{figure}[hbt!]
  \includegraphics[scale=.8]{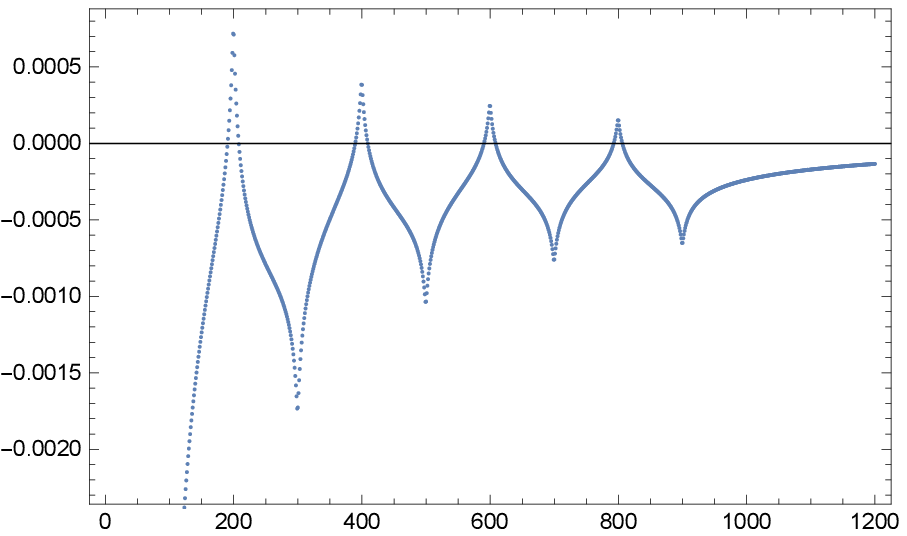}
  \caption{Gravitational field of the homogeneous five ring model with $1/R$ density}
 \end{figure}
 \subsection{Exponentially decreasing density of mass}
As a next example, we present a case where the density decreases exponentially.    
 
$\phi(r)= -\frac{1}{100}(\text{NIntegrate}\left[\frac{R \exp \left(-\left(\frac{R}{500}\right)^2\right)}{\sqrt{r^2-2 r R \cos (y)+R^2+z^2}},\{y,0,2 \pi \},\{z,0.05,5\},\{R,0,100\}\right]\\ 
+\text{NIntegrate}\left[\frac{R \exp \left(-\left(\frac{R}{500}\right)^2\right)}{\sqrt{r^2-2 r R \cos (y)+R^2+z^2}},\{y,0,2 \pi \},\{z,0.05,5\},\{R,200,300\}\right]\\
 +\text{NIntegrate}\left[\frac{R \exp \left(-\left(\frac{R}{500}\right)^2\right)}{\sqrt{r^2-2 r R \cos (y)+R^2+z^2}},\{y,0,2 \pi \},\{z,0.05,5\},\{R,400,500\}\right]\\
 +\text{NIntegrate}\left[\frac{R \exp \left(-\left(\frac{R}{500}\right)^2\right)}{\sqrt{r^2-2 r R \cos (y)+R^2+z^2}},\{y,0,2 \pi \},\{z,0.05,5\},\{R,600,700\}\right]\\
 +\text{NIntegrate}\left[\frac{R \exp \left(-\left(\frac{R}{500}\right)^2\right)}{\sqrt{r^2-2 r R \cos (y)+R^2+z^2}},\{y,0,2 \pi \},\{z,0.05,5\},\{R,800,900\}\right]).$
 
 The course of the potential is shown in Fig. 9.

 \begin{figure}[hbt!]
  \includegraphics[scale=.8]{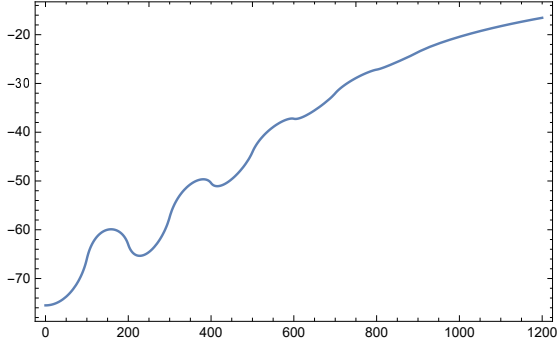}
  \caption{Gravitational potential $\phi (r)$ of the five rings model with exponentially decreasing density}
 \end{figure}

 \subsection{Decreasing high of homogeneous rings}

Similarly, we express the course of potential and field for a model of homogeneous rings of different high and the same width. The high decrease inversely proper with distance from the center. The results are in Fig. 11 and Fig. 12.
\begin{figure}[hbt!]
\includegraphics[scale=.8]{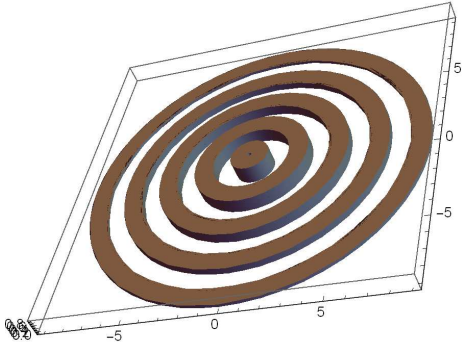}
\caption{Decreasing high of rings}
\end{figure}
 
 $
\phi(r)=\frac{1}{100} (\text{NIntegrate}[\frac{R}{\sqrt{r^2-2 r R \cos (y)+R^2+(100-0.1 R)^2}},\{y,0,2 \pi \},\{R,0,100\}]\\+\text{NIntegrate}[\frac{R}{\sqrt{r^2-2 r R \cos (y)+R^2+(100-0.1 R)^2}},\{y,0,2 \pi \},\{R,200,300\}]\\+\text{NIntegrate}[\frac{R}{\sqrt{r^2-2 r R \cos (y)+R^2+(100-0.1 R)^2}},\{y,0,2 \pi \},\{R,400,500\}]\\+\text{NIntegrate}[\frac{R}{\sqrt{r^2-2 r R \cos (y)+R^2+(100-0.1 R)^2}},\{y,0,2 \pi \},\{R,600,700\}]\\+\text{NIntegrate}[\frac{R}{\sqrt{r^2-2 r R \cos (y)+R^2+(100-0.1 R)^2}},\{y,0,2 \pi \},\{R,800,900\}])$

 \begin{figure}[hbt!]
  \includegraphics[scale=.8]{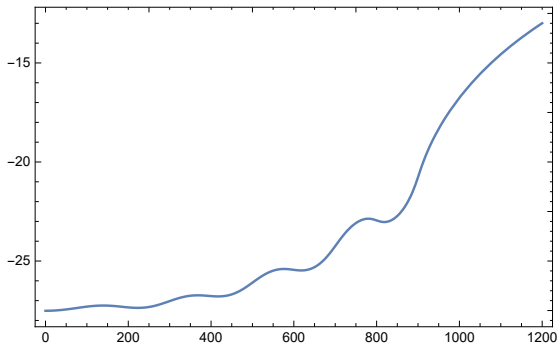}
  \caption{Gravitational potential $\phi (r)$ of the five rings model with decreasing high}
 \end{figure}
 \begin{figure}
  \includegraphics[scale=.8]{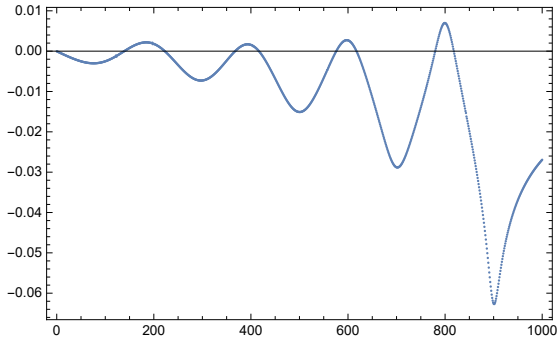}
  \caption{Gravitational field of the five rings model with decreasing high}
 \end{figure} 
  
 \section{Conclusion}

Due to the course of the force, another possibility is the superposition of the orbital motion with an oscillating radial motion around the circle, where the force of the ring itself is zero.
All results show a similar course of potentials. This contribution to the force caused by the central objects will affect the attractive force in the given places of the individual rings. The orbital velocity for a stable orbit is then given by the superposition of these forces. 

For simplicity, we calculated the field in the plane of symmetry. If we calculate the potential outside the plane $ z = 0, the $ potential of the models will have the same course, see the computation commands in Mathematics. However, the gradient will be different due to the loss of symmetry .

If the stars are closer to the center of the galaxy on the front of the ring and the stars further from the center of the galaxy on the outside of another ring, then it is possible that the orbital velocities of stars farther from the center are greater than the orbital speeds of the stars closer to the center of the galaxy.

\section*{Acknowledgement}

Author acknowledges the financial support from RVO14000, Ministry of Education, Youth and Sports.


\begin{thebibliography}{9}

\bibitem{Dark} https:$/ /en.wikipedia.org/wiki/Dark\,\,matter$

\bibitem{B3} https: $/ /en.wikipedia.org/wiki/List\, of\, ring\, galaxies $

\bibitem{B4}  https:$//en.wikipedia.org/wiki/Milky\,Way$

\end{thebibliography}
\end{document}